\documentclass[pra,twocolumn,showpacs]{revtex4}

\usepackage{graphicx}
\usepackage{textcomp}
\usepackage{amsmath}
\usepackage{inputenc}

\newcommand{\bra}[1]{\ensuremath{\left\langle\,#1\,\right|}}                              
\newcommand{\ket}[1]{\ensuremath{\left|\,#1\,\right\rangle}}

\newcommand{\pe}{\mbox{$^\prime$}}
\newcommand{\tabentry}[3]{\ #1&$\pm$&\!\!#2&\!#3}
\newcommand{\ctabentry}[2]{\multicolumn{3}{|c}{#1}&\!\!\!\!#2}
\newcommand{\rtabentry}[2]{\multicolumn{3}{|r}{#1}&\!\!\!\!#2}

\begin{document}
\date{May 2, 2005}

\author{S.~Kuhr}
\email{kuhr@lkb.ens.fr}
\author{W.~Alt}
\author{D.~Schrader}
\author{I.~Dotsenko}
\author{Y.~Miroshnychenko}
\author{A.~Rauschenbeutel}
\author{D.~Meschede}
\affiliation{Institut f\"ur Angewandte Physik, Universit\"at
Bonn, Wegelerstr.~8, D-53115 Bonn, Germany}
\title{Analysis of dephasing mechanisms in a standing wave dipole trap}

\begin{abstract}
We study in detail the mechanisms causing dephasing of hyperfine
coherences of cesium atoms confined by a far off-resonant standing
wave optical dipole trap [S. Kuhr {\it et al.}, Phys. Rev. Lett.
{\bf 91}, 213002 (2003)]. Using Ramsey spectroscopy and spin echo
techniques, we measure the reversible and irreversible dephasing
times of the ground state coherences. We present an analytical
model to interpret the experimental data and identify the
homogeneous and inhomogeneous dephasing mechanisms. Our scheme to
prepare and detect the atomic hyperfine state is applied at the
level of a single atom as well as for ensembles of up to 50 atoms.
\end{abstract}

\pacs{32.80.Lg, 32.80.Pj, 42.50.Vk}

\maketitle
\section{Introduction}

The coherent manipulation of isolated quantum systems has received
increased attention in the recent years, especially due to its
importance in the field of quantum computing. A possible quantum
computer relies on the coherent manipulation of quantum bits
(qubits), in which information is also encoded in the quantum
phases. The coherence time of the quantum state superpositions is
therefore a crucial parameter to judge the usefulness of a system
for storage and manipulation of quantum information. Moreover,
long coherence times are of great importance for applications in
precision spectroscopy such as atomic clocks.

Information cannot be lost in a closed quantum system since its
evolution is unitary and thus reversible. However, a quantum
system can never be perfectly isolated from its environment. It is
thus to some extent an open quantum system, characterized by the
coupling to the environment \cite{Zurek82}. This coupling causes
decoherence, i.\,e.~the evolution of a pure quantum state into a
statistical mixture of states. Decoherence constitutes the
boundary between quantum and classical physics \cite{Zurek91}, as
demonstrated in experiments in Paris and Boulder
\cite{Brune96,Haroche98,Monroe96}. There, decoherence was observed
as the decay of macroscopic superposition states (Schr\"odinger
cats) to statistical mixtures.

We can distinguish decoherence due to the progressive entanglement
with the environment from dephasing effects caused by classical
fluctuations. This dephasing of quantum states of trapped
particles has recently been studied both with ions
\cite{Schmidt-Kaler03} and neutral atoms in optical traps
\cite{Davidson95,Ozeri99}. In this work, we have analyzed
measurements of the dephasing mechanisms acting on the hyperfine
ground states of cesium atoms in a standing wave dipole trap. More
specifically, we use the two Zeeman sublevels \ket{F=4,m_F=0} and
\ket{F=3,m_F=0} which are coupled by microwave radiation at
$\omega_{\rm hfs}/2\pi=9.2$~GHz.

We present our setup and the relevant experimental tools in
Sec.~\ref{sec:setup}, with special regard to the coherent
manipulation of single neutral atoms.  Our formalism and the
notation of the dephasing/decoherence times are briefly introduced
in Sec.~\ref{sec:definition}. Finally, in Secs.~\ref{sec:ramsey}
and~\ref{sec:echo} we experimentally and theoretically analyze the
inhomogeneous and homogeneous dephasing effects.

\section{Experimental tools}\label{sec:setup}
\subsection{Setup}
\begin{figure}[!b]
\begin{center}
  \includegraphics[width=8.5cm]{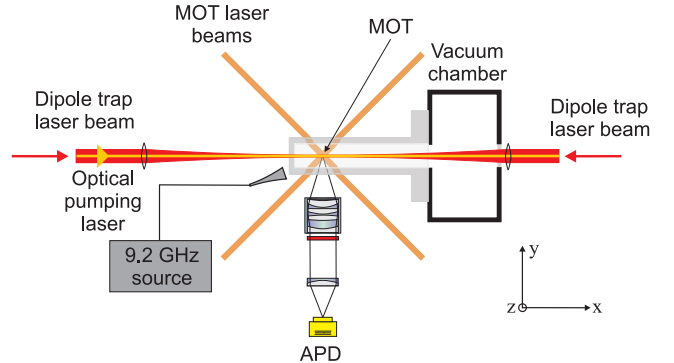}
\end{center}
\vspace{-0.5cm} \caption{Experimental setup.} \label{fig:setup}
\end{figure}

We trap and manipulate cesium atoms in a red detuned standing wave
dipole trap. Our trap is formed of two counterpropagating Gaussian
laser beams with waist $2w_0=40$~\textmu m and a power of max.~2~W
per beam (see Fig.~\ref{fig:setup}), derived from a single Nd:YAG
laser ($\lambda=1064$~nm). Typical trap depths are on the order of
$U_0=1$~mK. The laser beams have parallel linear polarization and
thus produce a standing wave interference pattern. Both laser
beams are sent through acousto-optic modulators (AOMs), to
mutually detune them for the realization of a moving standing
wave. This ``optical conveyor-belt'' was introduced in previous
experiments \cite{Kuhr01,Schrader01} and has been used for the
demonstration of quantum state transportation \cite{Kuhr03}. For
the experiments in this paper, however, we do not transport the
trapped atoms. To eliminate any heating effect arising due to the
phase noise of the AOM drivers \cite{Schrader01,Alt02b}, we used
the non-deflected beams (0$^{\rm th}$~order of the AOMs) to form
the dipole trap. The AOMs are only used to vary the dipole trap
laser intensity by removing power from the trap laser beams.

Cold atoms are loaded into the dipole trap from a high gradient
magneto-optical trap (MOT). The high field gradient of the MOT
($\partial B/\partial z=340$~G/cm) is produced by water cooled
magnetic coils, placed at a distance of 2~cm away from the trap.
The magnetic field can be switched to zero within 60~ms (limited
by eddy currents in the conducting materials surrounding the
vacuum chamber) and it can be switched back on within 30~ms. Our
vacuum chamber consists of a glass cell, with the cesium reservoir
being separated from the main chamber by a valve. Cesium atoms are
loaded into the MOT at random from the background gas vapor. To
speed up the loading process, we temporarily lower the magnetic
field gradient to $\partial B/\partial z=25$~G/cm during a time
$t_{\rm low}$. The low field gradient results in a larger capture
cross section which significantly increases the loading rate.
Then, the field gradient is returned to its initial value,
confining the trapped atoms at the center of the MOT. Varying
$t_{\rm low}$ enables us to select a specific average atom number
ranging from 1 to 50. The required time depends on the cesium
partial pressure, which was kept at a level such that we load
typically 50 atoms within $t_{\rm low}=100$~ms in these
experiments.

In order to transfer cold atoms from the MOT into the dipole trap,
both traps are simultaneously operated for some tens of
milliseconds before we switch off the MOT. After an experiment in
the dipole trap the atoms are transferred back into the MOT by the
reverse procedure. All our measurements rely on counting the
number of atoms in the MOT before and after any intermediate
experiment in the dipole trap. For this purpose we collect their
fluorescence light by a home-built diffraction-limited objective
\cite{Alt02a} and detect the photons with an avalanche photodiode
(APD).

Three diode lasers are employed in this experiment which are set
up in Littrow configuration and locked by polarization
spectroscopy. The MOT cooling laser is stabilized to the
$F=4\rightarrow F\pe=3/F\pe=5$ crossover transition and shifted by
an AOM to the red side of the cooling transition $F=4\rightarrow
F\pe=5$. The MOT repumping laser is locked to the $F=3\rightarrow
F\pe=4$ transition, it is $\pi$-polarized and is shined in along
the dipole trap axis. To optically pump the atoms into the
$\ket{F=4, m_F=0}$ state, we use the unshifted MOT cooling laser,
which is only detuned by +25~MHz from the required $F=4\rightarrow
F\pe=4$ transition. This small detuning is partly compensated for
by the light shift of the dipole trap. We shine in the laser along
the dipole trap axis with $\pi$-polarization together with the MOT
repumper. We found that 80\% of the atoms are pumped into the
$\ket{F=4, m_F=0}$ state, presumably limited by polarization
imperfections of the optical pumping lasers.

For the state selective detection (see below) we use a
``push-out'' laser, resonant to the $F=4\rightarrow F\pe=5$
transition. It is $\sigma^+$-polarized and shined in perpendicular
to the trapping beams (z-axis in Fig.~\ref{fig:setup}).

To generate microwave pulses at the frequency of 9.2~GHz we use a
synthesizer (Agilent 83751A), which is locked to an external
rubidium frequency standard (Stanford Research Systems, PRS10).
The amplified signal ($P=+36~\mbox{dBm}=4.0$~W) is radiated by a
dipole antenna, placed at a distance 5~cm away from the MOT.

Compensation of the earth's magnetic field and stray fields
created by magnetized objects close to the vacuum cell is achieved
with three orthogonal pairs of coils. For the compensation, we
minimize the Zeeman splitting of the hyperfine ground state $m_F$
manifold which is probed by microwave spectroscopy. Using this
method we achieve residual fields of $B_{\rm res} < 0.4$~\textmu T
(4~mG). The coils of the $z$-axis also serve to produce a guiding
field, which defines the quantization axis.

\subsection{State selective detection of a single neutral atom}
Sensitive experimental methods had to be developed in order to
prepare and to detect the atomic hyperfine state at the level of a
single atom. State selective detection is performed by a laser
which is resonant with the $F=4 \rightarrow F'=5$ transition and
thus pushes the atom out of the dipole trap if and only if it is
in $F=4$. An atom in the $F=3$ state, however, is not influenced
by this laser. Thus, it can be transferred back into the MOT and
be detected there. Although this method appears complicated at
first, it is universal, since it works with many atoms as well as
with a single one. Other methods, such as detecting fluorescence
photons in the dipole trap by illuminating the atom with a laser
resonant to the $F=4 \rightarrow F\pe=5$ transition, failed in our
case because the number of photons detected before the atom decays
into the $F=3$ state is not sufficient.

In order to achieve a high efficiency of the state selective
detection process, it is essential to remove the atom out of the
dipole trap before it is off-resonantly excited to $F\pe=4$ and
spontaneously decays into the $F=3$ state. For this purpose, we
use a $\sigma^{+}$-polarized push-out laser, such that the atom is
optically pumped into the cycling transition $\ket{F=4, m_F=4}
\rightarrow \ket{F\pe=5, m_F=5}$. In our setup, the polarization
is not perfectly circular, since for technical reasons we had to
shine in the laser beam at an angle of $2^\circ$ with respect to
the magnetic field axis. This entails an increased probability of
exciting the $F'=4$ level from where the atom can decay into the
$F=3$ ground state. To prevent this, we remove the atom from the
trap sufficiently fast by shining in the push-out laser from the
radial direction with high intensity ($I/I_0\approx 100$, with
$w_0 = 100$~\textmu m, $P = 30$~\textmu W, where
$I_0=1.1$~mW/cm$^2$ is the saturation intensity). In this regime
its radiation pressure force is stronger than the dipole force in
the radial direction, such that we push out the atom within half
the radial oscillation period ($\approx~1$~ms). In this case, the
atom receives a momentum corresponding to the sum of all
individual photon momenta. This procedure is more efficient than
heating an atom out of the trap, which occurs when the radiation
pressure force of the push-out laser is weaker than the dipole
force, and the atom performs a random walk in momentum space while
absorbing and emitting photons.





If we adiabatically lower the trap to typically $0.12$~mK prior to
the application of the push-out laser, we need on average only
35 photons to push the atom out of the trap. This
number is small enough to prevent off-resonant excitation to
$F\pe=4$ and spontaneous decay to $F=3$.

\begin{figure}
\begin{center}
  \includegraphics[width=8.5cm]{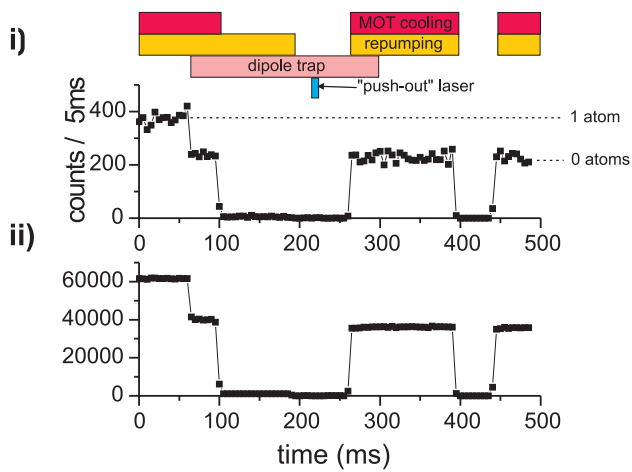}
\end{center}
\vspace{-0.3cm} \caption{State selective detection of a single
atom. The graphs show the fluorescence signal of the atom during
state preparation and detection, binned in time intervals of 5 ms.
The bars above the graphs show the timing of the lasers. Graphs
(a)(i) and (b)(i) show the signals of a single atom, prepared in
$F=3$ and $F=4$, respectively. Graphs (a)(ii) and (b)(ii) show the
added signal of about 150 events. } \label{fig:SSD}
\end{figure}

A typical experimental sequence to test the state selective
detection is shown in Fig.~\ref{fig:SSD}. First, the atom is
transferred from the MOT into the optical dipole trap. Using the
cooling and repumping laser of the MOT, we optically pump the atom
either in the $F=3$ (Fig.~\ref{fig:SSD}a) or the $F=4$ hyperfine
state (Fig.~\ref{fig:SSD}b). The push-out laser then removes all
atoms in $F=4$ from the trap. Any remaining atom in $F=3$ is
transferred back into the MOT, where it is detected.

Figures~\ref{fig:SSD}a(i) and \ref{fig:SSD}b(i) show the signals
of a single atom, prepared in $F=3$ and $F=4$, respectively. Our
signal-to-noise ratio enables us to unambiguously detect the
surviving atom in $F=3$, demonstrating the state-selective
detection at the single atom level. We performed 157 repetitions
with a single atom prepared in $F=3$ and found that in 153 of the
cases the atom remains trapped, yielding a detection probability
of $97.5^{+1.2}_{-2.0}\%$. Similarly, only 2 out of 167 atoms
prepared in $F=4$ remain trapped, yielding 1.2$^{+1.6}_{-0.8}\%$.
The asymmetric errors are the Clopper-Pearson 68\% confidence
limits \cite{Barlow}. These survival probabilities can also be
inferred by directly adding the signals of the individual
repetitions and by comparing the initial and final fluorescence
levels in the MOT, see Figs.~\ref{fig:SSD}a(ii) and
\ref{fig:SSD}b(ii).

\begin{figure}[t]
\begin{center}
  \includegraphics[width=8.5cm]{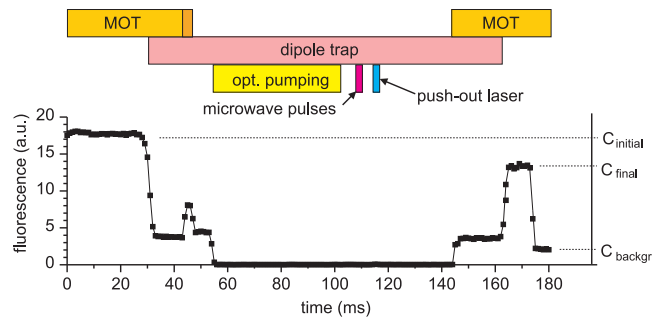}
\end{center}
\vspace{-0.5cm}
 \caption{Atom counting. Initial and final numbers of atoms
are inferred from their fluorescence in the MOT. Shown are the
integrated APD counts binned in time intervals of 1~ms and
accumulated over 10 repetitions with 20 atoms each. }
\label{fig:AtomCounting}
\end{figure}
All following experiments are performed in the same way. We
initially prepare the atoms in the \ket{F=4,m_F=0} state and
measure the population transfer to \ket{F=3, m_F=0} induced by
the microwave radiation. After the application of one or a
sequence of microwave pulses, the atom is in general in a
superposition of both hyperfine states,
\begin{equation}
    \ket{\psi} = c_3 \ket{F=3, m_F=0} + c_4 \ket{F=4, m_F=0},
\end{equation}
with complex probability amplitudes $c_3$ and $c_4$. Our detection
scheme only allows us to measure the population of the hyperfine
state $F=3$:
\begin{equation}\label{e:P3(w)}
  P_3 = |c_3|^2 = \frac{w+1}{2},
\end{equation}
where $w$ is the third component of the Bloch vector, see below.
The number $P_3$ is determined from the number of atoms before
($N_{\rm initial}$) and after ($N_{\rm final}$) any experimental
procedure in the dipole trap. $N_{\rm initial}$ and $N_{\rm
final}$ are inferred from the measured photon count rates, $C_{\rm
initial}$, $C_{\rm final}$ and $C_{\rm backgr}$ (see
Fig.~\ref{fig:AtomCounting}):
\begin{equation}\label{e:N_initial and N_final}
    N_{\rm initial} = \frac{C_{\rm initial}-C_{\rm backgr}}{C_{\rm 1atom}}
\end{equation}
and
\begin{equation}
    N_{\rm final} = \frac{C_{\rm final}-C_{\rm backgr}}{C_{\rm 1atom}}.
\end{equation}
The fluorescence rate of a single atom, $C_{\rm 1atom}$, is
measured independently. From the atom numbers we obtain the
fraction of atoms transferred to $F=3$,
\begin{equation}\label{e:P3}
    P_3 = \frac{N_{\rm final}}{N_{\rm initial}}
\end{equation}
The measured number of atoms, $N_{\rm inital}$ can be larger than
the actual number of atoms in the dipole trap, since we lose atoms
during the transfer from the MOT into the dipole trap (see below).



\subsection{Rabi oscillations}\label{ch:ExpRabiOsz}

\begin{figure}
\begin{center}
  \includegraphics[width=7cm]{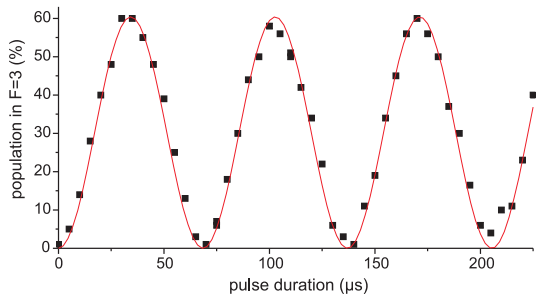}
\end{center}
\vspace{-0.5cm}\caption{Rabi oscillations on the $\ket{F=4,
m_F=0}\rightarrow \ket{F=3, m_F=0}$ clock transition recorded at a
trap depth $U_0=1.0$~mK. Each data point results from 100 shots
with about 60 initial atoms. The line is a fit according to
Eq.~(\ref{e:p3RabiOscillation}).} \label{fig:RabiOscillations}
\end{figure}

We induce Rabi oscillations by a single resonant microwave pulse
at the maximum RF power. For the graph of
Fig.~\ref{fig:RabiOscillations} we varied the pulse length from
0~\textmu s to 225~\textmu s in steps of 5~\textmu s. Each point
in the graph results from 100 shots with about $60\pm10$ atoms
each. The corresponding statistical error is below 1\% and is thus
not shown in the graph. The error of the data points in
Fig.~\ref{fig:RabiOscillations} is dominated by systematic drifts
of the storage probability and efficiencies of the state
preparation and detection. Since $w(t)=-\cos{\Omega_{\rm R}t}$, we
fit the graph with
\begin{equation}\label{e:p3RabiOscillation}
  P_{\rm 3} (t) = \dfrac{C}{2} \left(1 - \cos \Omega_{\rm R} t\right),
\end{equation}
which yields a Rabi frequency $\Omega_{\rm R}/2\pi=(14.60\pm
0.02)$~kHz. Note that this Rabi frequency is higher than the one
used later in this report (10~kHz) because we changed the position
of the microwave antenna for practical reasons. The maximum
population detected in $F=3$ is $C=(60.4\pm 0.7)$\%. This
reduction from 100\% is caused by two effects. First, when we use
many ($>40$) atoms at a time, up to 20\% of the atoms are lost
during the transfer from the MOT into the dipole trap due to
inelastic collisions, as verified in an independent measurement.
The remaining losses arise due to the non-perfect optical pumping
process.

\section{Phenomenological description of decoherence and
dephasing}\label{sec:definition}

In our experiment we observe quantum states in an ensemble
average, and decoherence manifests as a decay or dephasing of the
induced magnetic dipole moments. It is useful to distinguish
between homogeneous and inhomogeneous effects. Whereas homogeneous
dephasing mechanisms affect each atom in the same way,
inhomogeneous dephasing only appears when observing an ensemble of
many atoms possessing slightly different resonance frequencies. As
we will see later, the most important difference between the two
mechanisms is the fact that inhomogeneous dephasing can be
reversed, in contrast to the irreversible homogeneous dephasing.

The interaction between the oscillating magnetic field component
of the microwave radiation, $B \cos{\omega t}$, and the magnetic
dipole moment, $\mu$, of the atom is well approximated by the
optical Bloch equations \cite{AllenEberly}:
\begin{equation}\label{e:BlochVector}
  \dot{\boldsymbol u}= - \boldsymbol \Omega \times \boldsymbol u
\end{equation}
with the torque vector $\boldsymbol \Omega \equiv (\Omega_{\rm R},
0, \delta)$ and the Bloch vector $\boldsymbol u \equiv (u,v,w)$.
Here, $\Omega_{\rm R}=\mu B/\hbar$ is the Rabi frequency and
$\delta=\omega-\omega_0$ is the detuning of the microwave from the
atomic transition frequency $\omega_0$. In the following, the
initial quantum state \ket{F = 4, m_F=0} corresponds to the Bloch
vector $\boldsymbol u=(0,0,-1)$, whereas \ket{F = 3, m_F=0}
corresponds to $\boldsymbol u=(0,0,1)$.

We include the decay rates as damping terms into the Bloch
equations and use a notation of the different times for population
and polarization decay similar to the one of nuclear magnetic
resonance:
\begin{subequations}
\begin{eqnarray}\label{e:BlochDampingExpt}
\dot{\left<u\right>} & =& \delta \left<v\right> - \frac{\left<u\right>}{T_2}\\
\dot{\left<v\right>} & =& -\delta \left<u\right> + \Omega_{\rm R}
\left<w\right>
- \frac{\left<v\right>}{T_2}\\
\dot{\left<w\right>} & =&- \Omega_{\rm R} \left<v\right>-
\frac{\left<w\right>-w_{\rm st}}{T_1},
\end{eqnarray}
\end{subequations} where $\left<\dotsc\right>$ denotes the
ensemble average. The total homogeneous transverse decay time
$T_2$ is given by the polarization decay time $T_2\pe$ and the
reversible dephasing time $T_2^*$
\begin{equation}\label{e:tTimes}
    \frac{1}{T_2} = \frac{1}{T_2\pe} + \frac{1}{T_2^*}.
\end{equation}
Inhomogeneous dephasing ($T_2^*$) occurs because  the atoms may
have different resonance frequencies depending on their
environment. Thus the Bloch vectors of the individual atoms
precess with different angular velocities and lose their phase
relationship, they dephase. In our case, inhomogeneous dephasing
arises due to the energy distribution of the atoms in the trap.
This results in a corresponding distribution of light shifts
because hot and cold atoms experience different average trapping
laser intensities.

The longitudinal relaxation time, $T_1$, describes the population
decay to a stationary value $w_{\rm st}$. In our case, $T_1$ is
governed by the scattering of photons from the dipole trap laser,
which couples the two hyperfine ground states via a two-photon
Raman transition. This effect is suppressed due to a destructive
interference effect yielding relaxation times of several seconds
(see Sec.~\ref{subsec:originsOfDephasing}). We do not include
losses of atoms from the trap in the decay constants, which occur
on the same timescale.

\section{Inhomogeneous dephasing}\label{sec:ramsey}
We measure the transverse decay time $T_2$ by performing Ramsey
spectroscopy, which consists of the application of two coherent
rectangular microwave pulses, separated by a time interval $t$
\cite{Ramsey}. The initial Bloch vector $\boldsymbol u_0 =
(0,0,-1)$ corresponds to an atom prepared in the \ket{F=4,
m_F=0}-state. A $\pi/2$-pulse rotates the Bloch vector into the
state $(0,-1,0)$, where the atom is in a superposition of both
hyperfine states. The Bloch vector freely precesses in the
$uv$-plane with an angular frequency
$\delta$. 
Note that $\delta$ has to be small compared to the Rabi frequency
and the spectral pulse width, such that the pulse can be
approximated as near resonant, and complete population transfer
can occur. After a free precession during $t$, a second
$\pi/2$-pulse is applied. The measurement of the quantum state
finally projects the Bloch vector onto the $w$ axis.

\begin{figure}
\begin{center}
  \includegraphics[width=8cm]{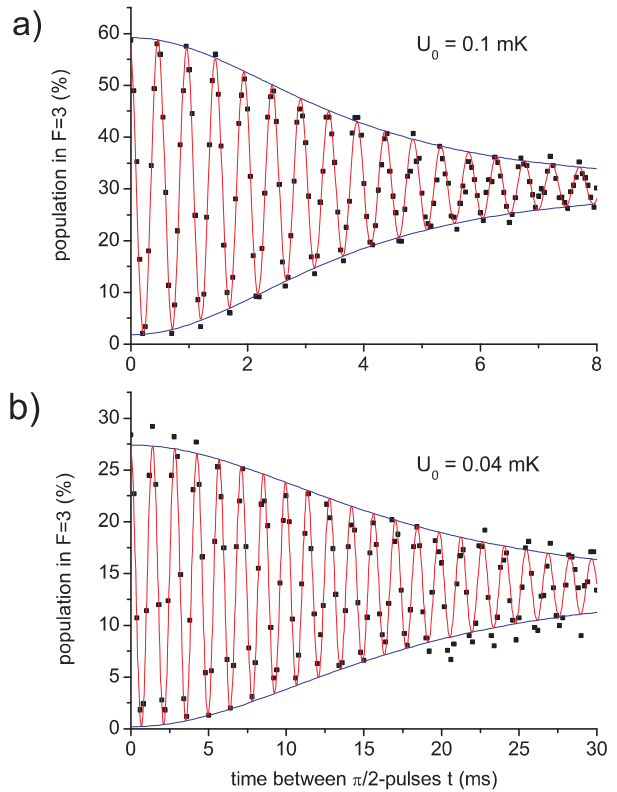}
\end{center}
\vspace{-0.5cm}\caption{Ramsey fringes recorded for two different
trap depths (a) $U_0=0.1$~mK and (b) $0.04$~mK. Their decay with
time constants $T_2^*=4.4\pm0.1$~ms and $20.4\pm1.1$~ms,
respectively, is governed by inhomogeneous dephasing caused by the
energy distribution in the trap. Each data point results from 30
shots with about 50 initial atoms. The damped oscillation is a fit
with $P_{\rm 3,Ramsey}(t)$ and the envelopes are the functions
$B\pm A\alpha(t,T_2^*)$ (see Eqs.~(\ref{e:alpha(t)}),
(\ref{e:FITp3Ramsey})). \label{fig:Ramsey_Experiment}}
\end{figure}

We recorded Ramsey fringes for two different dipole trap depths,
$U_0=0.1$~mK and $0.04$~mK (see Fig.~\ref{fig:Ramsey_Experiment}).
Each point in the graph of Fig.~\ref{fig:Ramsey_Experiment}
corresponds to 30 shots with about 50 trapped atoms per shot,
yielding errors (not shown) of less than 1\%. The quoted values
for $U_0$ are calculated from the measured power and the waist of
the dipole trap laser beam and have an estimated uncertainty of up
to 50\%. We initially transfer the atoms from the MOT into a
deeper trap ($U_0>1$~mK) to achieve a high transfer efficiency.
When the MOT is switched off, we adiabatically lower the trap
depth using the AOMs.

Our Ramsey fringes show a characteristic decay, which is not
exponential. This decay is due to inhomogeneous dephasing, which
occurs because after the first $\pi/2$-pulse, the atomic
pseudo-spins precess with different angular frequencies. In the
following, we derive analytic expressions for the observed Ramsey
signal and we show that the envelope of the graphs of
Fig.~\ref{fig:Ramsey_Experiment} is simply the Fourier transform
of the atomic energy distribution.\\

\subsection{Differential light shift and decay of Ramsey fringes}

The light shift of the ground state due to the Nd:YAG laser is
simply the trapping potential
\begin{equation}\label{e:U02}
    U_0(\Delta) = \frac{\hbar\Gamma}{8} \frac{I}{I_0}
    \frac{\Gamma}{\Delta}.
\end{equation}
The detuning of the Nd:YAG-laser from the D-line of an atom in
$F=4$ is 9.2~GHz less than for an atom in $F=3$. As a consequence,
the $F=4$ level experiences a slightly stronger light shift,
resulting in a shift of the $F=3\rightarrow F=4$ microwave
transition towards smaller resonance frequencies. This
differential light shift, $\delta_{0}$, can be approximated as
\begin{equation}\label{e:diffLSAnsatz}
  \hbar \delta_{\rm 0}  = U_0(\Delta_{\rm eff}) - U_0(\Delta_{\rm eff}+\omega_{\rm
  hfs}),
\end{equation}
where $\Delta_{\rm eff}= - 1.2\times 10^7~\Gamma$ is an effective
detuning, taking into account the weighted contributions of the
D$_1$ and D$_2$ lines \cite{Schrader01}. $\omega_{\rm
hfs}=2.0\times 10^3~\Gamma$ is the ground state hyperfine
splitting. Since $\omega_{\rm hfs}\ll\Delta_{\rm eff}$, we find
that the differential light shift is proportional to the total
light shift $U_0$,
\begin{equation}\label{e:differentialLS}
    \hbar \delta_{\rm 0} =  \eta U_0,
\end{equation}
with a scaling factor $\eta=\omega_{\rm hfs}/\Delta_{\rm eff} =
1.45\times10^{-4}$. For atoms trapped in the bottom of a potential
of $U_0=1$~mK, the differential light shift is $\delta_{0}=-
2\pi\times 3.0$~kHz.

In the semiclassical limit, i.~e.~neglecting the quantized motion
of the atom in the dipole trap potential, the free precession
phase accumulated by an atomic superposition state between the two
$\pi/2$-pulses depends on the average differential light shift
only. In the following, we calculate the expected Ramsey signal
using this semiclassical approach and obtain simple analytical
expressions. Furthermore, we verified the validity of the
presented model by performing a quantum mechanical density matrix
calculation (not presented here) which agrees to within one
percent with the semiclassical results. The small deviation can be
attributed to the occurrence of small oscillator quantum numbers
$n_{\rm osc} \simeq 5$ in the stiff direction of the trap.

Note that, strictly speaking, our model of a time-averaged
differential light shift is only correct if the atom carries out
an integer number of oscillation periods in the trap between the
two $\pi/2$-pulses. However, we have checked that the variable
phase accumulated during the remaining fraction of an oscillation
period does not cause a measurable reduction of the Ramsey fringe
contrast and can therefore be neglected.

Since a hot atom experiences a lower laser intensity than a cold
one, its averaged differential light shift is smaller. The energy
distribution of the atoms in the dipole trap obeys a
three-dimensional Boltzmann distribution with probability density
\cite{Metcalf,korrMaxwell}
 \begin{equation}\label{e:Boltzmann}
    p(E) = \frac{E^2}{2(k_{\rm B}T)^{3}}\, \exp{\left(-\frac{E}{k_{\rm
    B}T}\right)}.
\end{equation}
Here $E=E_{\rm kin}+U$ is the sum of kinetic and potential energy.
In a harmonic trap the virial theorem states that the average
potential energy is half the total energy, $U = E/2$. Thus, the
average differential light shift for an atom with energy $E$ is
given by:
\begin{equation}\label{e:deltaLS}
    \delta_{\rm ls}(E) =  \delta_0 +
    \frac{\eta E}{2\hbar}
\end{equation}
where $\delta_0<0$ is the maximum differential light shift. As a
consequence, the energy distribution $p(E)$ yields, except for a
factor and an offset, an identical distribution
$\tilde{\alpha}(\delta_{\rm ls})$ of differential light shifts
\cite{korrMaxwell}:
\begin{equation}\label{e:distrDiffLS}
\tilde{\alpha}(\delta_{\rm ls}) = \frac{K^3}{2}
                        (\delta_{\rm ls}-\delta_{\rm 0})^2
                        \exp{[-K(\delta_{\rm ls}-\delta_{\rm 0})]}
\end{equation}
with
\begin{equation}
 K = \frac{2\hbar}{\eta k_{\rm B}T}.
\end{equation}
Note that this distribution is only valid in the regime $k_{\rm
B}T\ll U_0$, since the virial theorem was applied for the case of
a harmonic potential.

To model the action of the Ramsey pulse sequence, we express the
solutions of Eq.~(\ref{e:BlochVector}) as rotation matrices acting
on the Bloch vector. The Ramsey sequence then reads
\begin{equation}\label{e:Ramsey}
  \boldsymbol u_{\rm Ramsey}(t) =
  \Theta_{\pi/2}
  \cdot \Phi_{\rm free}(\delta,t)
  \cdot \Theta_{\pi/2}
  \cdot \boldsymbol u_0,
\end{equation}
with the matrices describing the action of a $\pi/2$-pulse,
\begin{equation}\label{e:Rot_PiOver2}
\Theta_{\pi/2} = \begin{pmatrix}
   1 & 0 & 0 \\
   0 & 0 & 1 \\
   0 & -1 & 0\
 \end{pmatrix},
\end{equation}
and the free precession around the $w$-axis with angular
frequency $\delta$ during a time interval $t$,
\begin{equation}\label{e:rot_free}
\Phi_{\rm free}(\delta,t) =\begin{pmatrix}
  \ \,\,\, \cos \phi(\delta,t)  & \sin \phi(\delta,t) & 0 \\
   -\sin \phi(\delta,t)  & \cos \phi(\delta,t) & 0 \\
   0 & 0 & 1 \
 \end{pmatrix}.
\end{equation}
The total precession angle, $\phi(\delta, t)$, represents the
accumulated phase during the free evolution of the Bloch vector,
$\phi(t) = \int_0^t \delta(t') dt'$. The detuning $\delta (t)$ may
in general vary spatially and in time, depending on the energy
shifts of the atomic levels.

If the Bloch vector is initially in the state $\boldsymbol u_0 =
(0,0,-1)$, we obtain from Eq.~(\ref{e:Ramsey}):
\begin{equation}\label{e:w_Ramsey}
    w_{\rm Ramsey}(t) = \cos \delta t,
\end{equation}
where $\delta = \omega-\omega_0$ is the detuning of the microwave
radiation with frequency $\omega$ from the atomic resonance
$\omega_0$. In order to see the Ramsey fringes, we purposely shift
$\omega$ with respect to the ground state hyperfine splitting,
$\omega_{\rm hfs}$, by a small detuning, $\delta_{\rm synth}$, set
at the frequency synthesizer
\begin{equation}
    \omega = \omega_{\rm hfs}+\delta_{\rm synth}.
\end{equation}
The atomic resonance frequency $\omega_0$ is modified due to
external perturbations
\begin{equation}
   \omega_0=\omega_{\rm hfs}+\delta_{\rm ls} + \delta_{\rm B},
\end{equation}
where $\delta_{\rm ls}$ is the energy dependent differential light
shift, $\delta_{\rm B}$ is the quadratic Zeeman shift.

Now, the inhomogeneously broadened Ramsey signal is obtained by
averaging over all differential light shifts $\delta_{\rm ls}$:
\begin{equation}\label{e:Ramseydeph}
\begin{array}{ll}
\!\!\! \!\!\! w_{\rm Ramsey, inh}(t) = & \displaystyle\!\!\!
 \int^{ \infty}_{\delta_{\rm 0}} \!\!\!
 \tilde{\alpha}(\delta_{\rm ls})\\
&  \!\!\!  \times\cos\left[(\delta_{\rm synth} - \delta_{\rm B} -
\delta_{\rm ls})
    t\right] d\delta_{\rm ls}.
\end{array}
\vspace{0.2cm}
\end{equation}
Eq.~(\ref{e:Ramseydeph}) shows that the shape of the Ramsey
fringes is the Fourier(-Cosine)-Transform of the atomic energy
distribution. Note that in the above integral we have set the
upper integration limit to $\infty$, instead of the maximum
physically reasonable value, $\delta_0/2$, to obtain the analytic
solution
\begin{equation}\label{e:w_analyt}
    w_{\rm Ramsey, inh}(t) = \alpha(t,T_2^*)\,
    \cos{\left[\delta\pe t + \kappa(t,T_2^*)\right]},
\end{equation}
with the sum of the detunings,
\begin{equation}
 \delta\pe = \delta_{\rm synth} - \delta_{\rm B}-\delta_{\rm 0}
\end{equation}
and a time- dependent amplitude $\alpha(t,T_2^*)$ and phase shift
$\kappa (t,T_2^*)$ \cite{korrMaxwell}
\begin{equation}\label{e:alpha(t)}
   \alpha(t,T_2^*) = \left[1+0.95\left(\frac{t}{T_2^*}\right)^2\right]^{-3/2}
\end{equation}
and
\begin{equation}\label{e:kappa(t)}
  \kappa (t,T_2^*) = -3\arctan\left(0.97 \frac{t}{T_2^*}\right).
\end{equation}

Despite this non-exponential decay, we have introduced the
inhomogeneous or reversible dephasing time $T_2^*$ as the 1/e-time
of the amplitude $\alpha(t)$:
\begin{equation}\label{e:def_T2*}
  T_2^* = \sqrt{e^{2/3}-1}\, K = 0.97 \frac{2\hbar}{\eta k_{\rm
  B}T}.
\end{equation}
Thus, the reversible dephasing time $T_2^*$ is inversely
proportional to the temperature of the atoms.

The phase shift  $\kappa(t,T_2^*)$ arises due to the asymmetry of
the probability distribution $\tilde{\alpha}(\delta_{\rm ls})$.
The hot atoms in the tail of the energy distribution dephase
faster than the cold atoms, due to their larger spread. The fact
that these hot atoms no longer contribute to the Ramsey signal
results in a weighting of the mean $\delta_{\rm ls}$ towards
larger negative values.

To fit our experimental data, we derive the following expression
from Eq.~(\ref{e:w_analyt}),
\begin{eqnarray}\label{e:FITp3Ramsey}
  P_{\rm 3,Ramsey} (t)&&=B + \alpha(t,T_2^*)\nonumber\\
&& \times A\cos{\left[ \delta\pe t + \kappa(t,T_2^*) +
\varphi\right]} ,
 \end{eqnarray}
where the amplitude $A$ and the offset $B$ account for the
imperfections of state preparation and detection. The other fit
parameters are $\delta\pe$, $T_2^*$, and a phase offset $\varphi$.

\begin{table}[!t]
\begin{center}
\begin{tabular}{|c|rcrl|rcrl|}
  \cline{2-9}
  \multicolumn{1}{c|}{}
  & \multicolumn{4}{|c|}{\rule[-2mm]{0mm}{6mm}Fig.~\ref{fig:Ramsey_Experiment}(a)}
  & \multicolumn{4}{|c|}{Fig.~\ref{fig:Ramsey_Experiment}(b)}
 \\
\hline
  $U_0 \mbox{ (est.)}$
  & \rtabentry{0.1 mK}{}
  & \rtabentry{0.04 mK}{} \\
  $\delta_{\rm synth}/2\pi$
  & \rtabentry{2250~Hz}{}
  & \rtabentry{1050~Hz}{} \\
%
  \hline
 $A$
  & \tabentry{28.7}{0.5}{\%}
  & \tabentry{13.6}{0.1}{\%}\\
  $B$
  & \tabentry{30.5}{0.1}{\%} 
  & \tabentry{13.8}{0.1}{\%} \\
  $\delta\pe/2\pi$
  & \tabentry{2133.7}{1.5}{Hz}
  & \tabentry{722.5}{0.5}{Hz}\\
  $\varphi$
  & \tabentry{0.35}{0.02}{}
  & \tabentry{0.13}{0.03}{} \\
  $T_2^*$
  & \tabentry{4.4}{0.1}{ms}
  & \tabentry{20.4}{0.6}{ms} \\
\hline
 \end{tabular}
\end{center}
\vspace{-0.5cm}\caption{Fit parameters extracted from the Ramsey
fringes of Fig.~\ref{fig:Ramsey_Experiment} using
Eq.~(\ref{e:FITp3Ramsey}).\label{tab:FitRamsey}}
\end{table}

The corresponding fits are shown in
Fig.~\ref{fig:Ramsey_Experiment}  and the resulting fit parameters
are summarized in Table~\ref{tab:FitRamsey}. For the two graphs,
the maximum population detected in $F=3$, $P_{3,{\rm max}}=A+B$ is
only about 60\% and 30\%, respectively. The reduction to 60\% in
Fig.~\ref{fig:Ramsey_Experiment}(a) is again due to imperfections
in the optical pumping process and due to losses by inelastic
collisions, as discussed in Sec.~\ref{ch:ExpRabiOsz}. The
additional reduction in Fig.~\ref{fig:Ramsey_Experiment}(b) occurs
during the lowering of the trap to $U_0=0.04$~mK, where another
50\% of the atoms are lost. Note, however, that the fringe
visibility
\begin{equation}\label{e:DefVisibility}
 V =\frac{A}{B}
\end{equation}
is not impaired by these imperfections. From the fit parameters we
obtain $V=0.97\pm0.01$ and $V=1.00^{+0}_{-0.03}$
for the two cases.

As a check of consistency, we calculate the differential light
shift $\delta_0$ from the fitted detuning $\delta\pe$ and the
experimental values of $\delta_{\rm B}$ and $\delta_{\rm synth}$,
\begin{equation}\label{e:delta0FromFit}
\delta_0=\delta_{\rm synth}-\delta_{\rm B}-\delta\pe.
\end{equation}
The calculated quadratic Zeeman shift in the externally applied
guiding field of $B=97.9\pm1.5$~\textmu T is $\delta_{\rm
B}/2\pi=412\pm13$~Hz, where the error is due to the uncertainty of
the calibration. We obtain $\delta_0/2\pi=-268\pm13$~Hz and
$\delta_0/2\pi=-78\pm13$~Hz. From the values of $\delta_0$ we can
formally deduce the potential depth corresponding to
$U_0=0.090\pm0.004$~mK and $U_0=0.026\pm0.004$~mK, which almost
match the expected trap depths estimated from the dipole trap
laser power assuming purely linear polarization. The discrepancy
for the lowest trap depth could arise from the fact that the
energy distribution is  truncated at $E = U_0$, since we have lost
the atoms with the highest energy during the lowering of the trap.
This truncation will reduce the effective $\delta_0$ and thus
yield a smaller trap depth.

%

Finally, the phase offset $\varphi$ occurs because the Bloch
vector precesses around the $w$-axis even during the application
of the two $\pi/2$-pulses. In contrast, our ansatz of
Eq.~(\ref{e:Ramsey}) takes into account only the free precession
in between the two pulses. The additional precession angle amounts
to $\varphi = 2\,t_{\pi/2}\,\delta\pe$. Given
$t_{\pi/2}=16$~\textmu s and the fitted value of $\delta\pe$ we
obtain $\varphi=0.42$ for Fig.~\ref{fig:Ramsey_Experiment}(a) and
$\varphi=0.14$ for Fig.~\ref{fig:Ramsey_Experiment}(b), which is
close to the fitted values of Table~\ref{tab:FitRamsey}.

\begin{figure*}[!t]
\begin{center}
  \includegraphics[width=18cm]{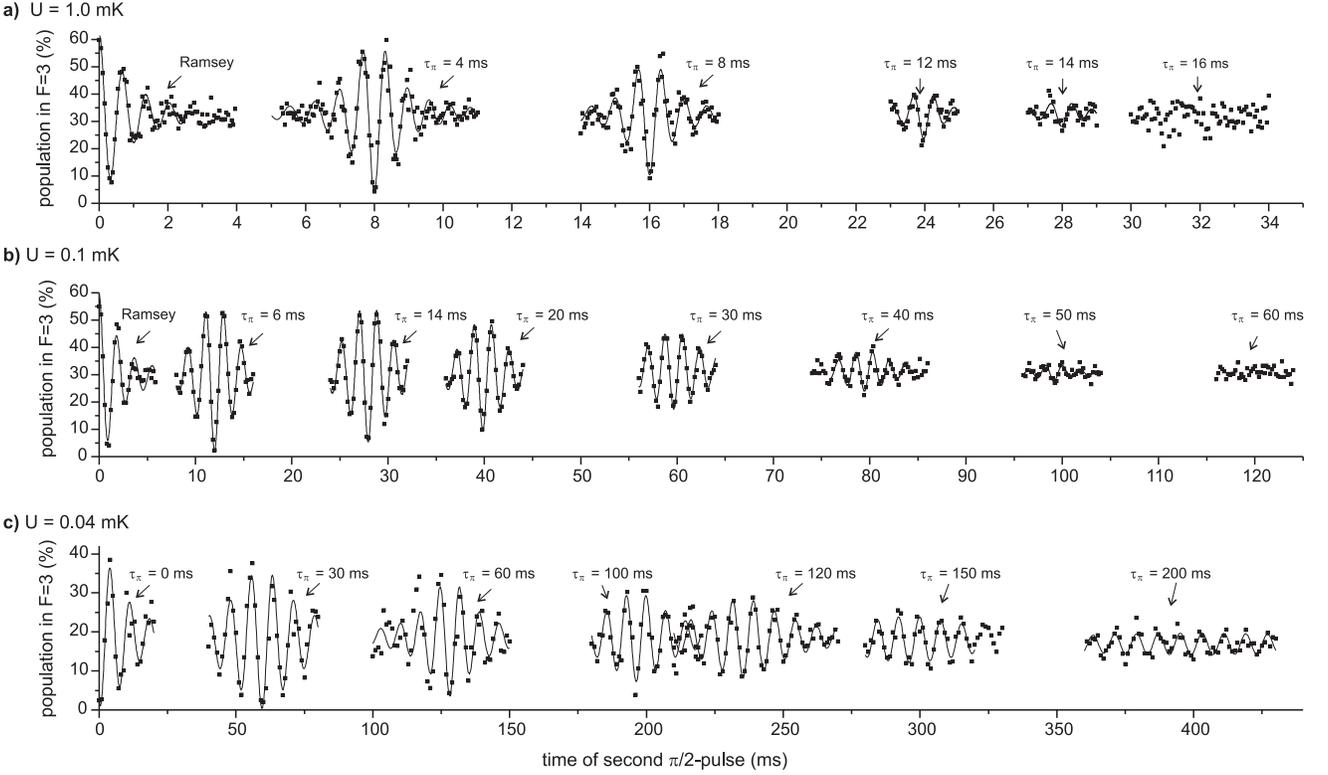}
\end{center}
\caption{Spin echoes. Shown are spin echoes recorded for three
different trap depths, (a) $U_0=1.0$~mK, (b) $0.1$~mK and (c)
$0.04$~mK. We observe a decrease of the maximum spin echo
amplitude with increasing time of the $\pi$-pulse, with longer
decay times in lower trap depths. All spin echoes are fitted using
Eq.~(\ref{e:FITp3Echo}). In (a) and (b), the first curve is a
Ramsey signal, recorded with otherwise identical parameters.
\label{fig:LotsOfEchoes}}
\end{figure*}

\section{Homogeneous dephasing mechanisms}\label{sec:echo}
\subsection{Spin echoes}

The inhomogeneous dephasing can be reversed using a spin echo
technique, i.~e.~by applying an additional $\pi$-pulse between the
two Ramsey $\pi/2$-pulses. Although originally invented in the
field of nuclear magnetic resonance \cite{Hahn50}, this technique
was recently also employed in optical dipole traps
\cite{Andersen03}.

We recorded echo signals in three different trap depths,
$U_0=1.0$~mK, $0.1$~mK and $0.04$~mK for different times of the
$\pi$-pulse, $\tau_\pi$ (see Fig.~\ref{fig:LotsOfEchoes}). We
observe that the visibility of the echo signals decreases if we
increase $\tau_\pi$. A slower decrease of the visibility is
obtained in lower traps. For $U_0=0.04$~mK, $\tau_\pi=200$~ms, we
even observed oscillations that reappear at $t=400$~ms.

\begin{figure}
  \centering
\includegraphics[width=8.6cm]{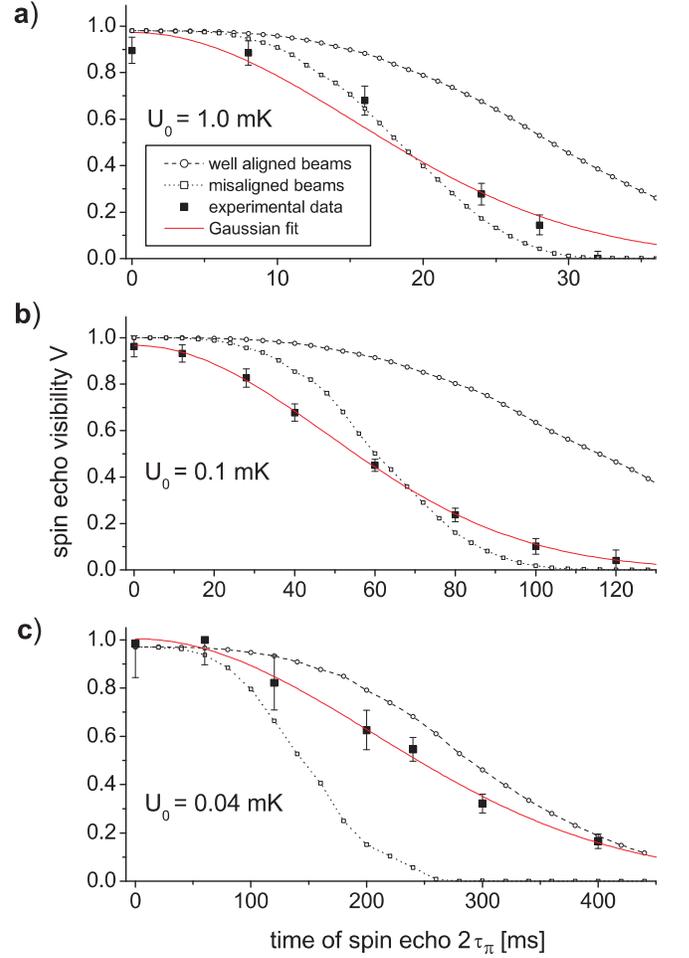}
\vspace{-0.5cm}\caption{Decay of the spin echo visibility,
extracted from the signals of Fig.~\ref{fig:LotsOfEchoes}. The
fits (red lines) are the Gaussians of
Eq.~(\ref{e:FitContrastOfSpinEcho}). The dashed and dotted lines
are the best and worst case predictions inferred from the measured
pointing instability of the trapping laser shown in
Fig.~\ref{fig:PointingInstabilities}.
  \label{fig:ContrastOfSpinEcho}}
\end{figure}

In order to interpret these results, we first model the action of
the microwave pulses for the spin echo, similar to the discussion
in Sec.~\ref{sec:ramsey}. After the first $\pi/2$-pulse at $t=0$,
all Bloch vectors start at $\boldsymbol u(0)=(0,-1,0)$. Due to
inhomogeneous dephasing, the Bloch vectors rotate at slightly
different frequencies around the $w$-axis. A $\pi$-pulse at time
$\tau_\pi$ rotates the ensemble of Bloch vectors around the $u$
axis by $180^\circ$ and induces a complete rephasing at
$2\tau_\pi$ in the state $\boldsymbol u(2\tau_\pi)=(0,1,0)$. The
corresponding matrix equation reads
\begin{eqnarray}\label{e:SpinEcho_Matrix}
  \boldsymbol u_{\rm echo}(t) &=&
  \Theta_{\pi/2}
  \cdot \Phi_{\rm free}(\delta,t-\tau_\pi)
  \cdot  \Theta_{\pi} \cdot  \nonumber \\
&&  \cdot  \Phi_{\rm free}(\delta,\tau_{\pi})
  \cdot  \Theta_{\pi/2}
  \cdot  \boldsymbol u_0,
\end{eqnarray}
where we defined
\begin{equation}\label{e:Rot_Pi}
\Theta_{\pi} = \begin{pmatrix}
   1 & 0 & 0 \\
   0 &-1 & 0 \\
   0 & 0 & -1\
 \end{pmatrix}.
\end{equation}
Here, $\tau_\pi$ is the time between the first $\pi/2$- and the
$\pi$-pulse, and $t>\tau_\pi$ is the time of the second $\pi/2$
pulse. As a result of Eq.~(\ref{e:SpinEcho_Matrix}), we obtain
\begin{equation}\label{e:w_echo}
    w_{\rm echo}(t) = - \cos[\delta (t-2\tau_{\pi})].
\end{equation}
We calculate the shape of the inhomogeneously broadened echo
signal, $w_{\rm echo, inh}(t)$, by integrating over all
differential light shifts $\delta_{\rm ls}$
\begin{eqnarray}\label{e:Echodeph_Int}
w_{\rm echo,inh}(t) &=&- \int^{\infty}_{\delta_{\rm 0}}\! \!\!\!
\tilde{\alpha}(\delta_{\rm ls})\times\nonumber\\
&&\hspace{-1.5cm}\times \cos \left[ (\delta_{\rm synth} -
\delta_{\rm ls} -\delta_{\rm B}) (t-2\tau_\pi)\right]
d\delta_{\rm ls}. \rule[-7mm]{7mm}{0mm}
\end{eqnarray}
The integration yields a result similar to
Eq.~(\ref{e:Ramseydeph}),
\begin{eqnarray}\label{e:echoAnalyt}
    w_{\rm echo, inh}(t) &=& - \alpha(t-2\tau_\pi)\nonumber\\
&&\!\!\times    \cos{\left[\delta\pe(t-2\tau_\pi) +
    \kappa(t-2\tau_\pi)\right]},
\end{eqnarray}
with amplitude $\alpha(t)$ and phase shift $\kappa(t)$ as defined
in Eqs.~(\ref{e:alpha(t)}) and (\ref{e:kappa(t)}). 
Eq.~(\ref{e:echoAnalyt}) shows that the amplitude of the echo
signal regains its maximum at time $2\tau_{\pi}$. Finally, the
population in $F=3$ reads:
\begin{eqnarray}\label{e:FITp3Echo}
P_{3, \rm echo}(t) &=& B-
 \alpha(t-2\tau_\pi,T_2^*)\nonumber\\
&&\hspace{-1.8cm} \times A\cos{\left[\delta\pe
(t\!-2\tau_\pi)+\kappa(t-2\tau_\pi,T_2^*)+\psi\right]}.
\end{eqnarray}
This equation is used to extract dephasing times $T_2^*$ from all
spin echoes of Fig.~\ref{fig:LotsOfEchoes}. The average values are
listed in Table~\ref{tab:times}, where $T_2^*$ was obtained by
averaging over the respective datasets. From the amplitude, $A$,
and offset, $B$, of each echo signal we calculate the visibility
$V=A/B$, plotted in Fig.~\ref{fig:ContrastOfSpinEcho} as a
function of $\tau_{\pi}$. The phase shift $\psi$ accounts for slow
systematic phase drifts during the spin echo sequence.

\begin{table}[!t]
\begin{center}
\begin{tabular}{|c|rcrl|rcrl|rcrl|}
  \cline{2-13}
  \multicolumn{1}{c|}{}
  & \multicolumn{4}{|c|}{\rule[-2mm]{0mm}{6mm}Fig.~\ref{fig:LotsOfEchoes}(a)}
  & \multicolumn{4}{|c|}{Fig.~\ref{fig:LotsOfEchoes}(b)}
  & \multicolumn{4}{|c|}{Fig.~\ref{fig:LotsOfEchoes}(c)}
 \\
\hline
  $U_0 \mbox{ (est.)}$
  & \rtabentry{1.0 mK}{}
  & \rtabentry{0.1 mK}{}
  & \rtabentry{0.04 mK}{} \\

  $T_2^*$
   & \tabentry{0.86}{0.05}{ms}
   & \tabentry{2.9}{0.1}{ms}
   & \tabentry{18.9}{1.7}{ms}\\

 $T_2\pe$
  & \tabentry{10.2}{0.4}{ms}
  & \tabentry{33.9}{1.0}{ms}
  & \tabentry{146.2}{6.6}{ms}\\

   $T_1$  (calc.)
  & \ctabentry{8.6~s}{}
  & \ctabentry{86~s}{}
  & \ctabentry{220~s}{}\\
\hline
 \end{tabular}
\end{center}
\vspace{-0.5cm}\caption{Summary of dephasing times. $T_2^*$ and
$T_2\pe$ are obtained from the echo signals of
Fig.~\ref{fig:LotsOfEchoes}. $T_1=\Gamma_{\rm Raman}^{-1}$ is
calculated using Eq.~(\ref{e:RamanVSRayleigh}). }\label{tab:times}
\end{table}

So far, we considered the detuning as constant during the
experimental sequence. We now include in our model a time-varying
detuning, $\delta(t)$, in order to account for a stochastic
variation of the precession angles of the Bloch vector,
\begin{equation}
\phi_1=\int_0^{\tau_\pi}\delta(t)\,dt \quad\mbox{and}\quad
\phi_2=\int_{\tau_\pi}^{2\tau_\pi}\delta(t)\,dt,
\end{equation}
before and after the $\pi$-pulse. The phase difference
$\phi_2-\phi_1$ is expressed as a mean difference of the detuning,
\begin{equation}
\Delta\delta = \frac{\phi_2-\phi_1}{\tau_\pi}.
\end{equation}
The Bloch vector at time $2\tau_\pi$, when the inhomogeneous
dephasing has been fully reversed, reads
\begin{eqnarray}\label{e:SpinEcho_Matrix_DeltaDelta}
  \boldsymbol u_{\rm echo}(\Delta\delta, 2\tau_\pi) &=&
  \Theta_{\pi/2}
  \cdot \Phi_{\rm free}(\delta+\Delta\delta,\tau_\pi)
  \cdot \Theta_{\pi}\cdot\nonumber\\
&&  \cdot \Phi_{\rm free}(\delta,\tau_{\pi})
  \cdot \Theta_{\pi/2}
  \cdot \boldsymbol u_0,
\end{eqnarray}
which results in
\begin{equation}\label{e:w(2TauPi)}
w_{\rm echo}(\Delta\delta,
2\tau_\pi)=-\cos(\Delta\delta\,\tau_\pi).
\end{equation}
For a Gaussian distribution of fluctuations with mean
$\overline{\Delta\delta}=0$ and variance $\sigma(\tau_\pi)^2$,
\begin{equation}\label{e:p(Deltadelta)}
p_{\tau_\pi}(\Delta\delta)=\frac{1}{\sigma(\tau_\pi)\sqrt{2\pi}}
\exp{\left[-\frac{(\Delta\delta)^2}{2\sigma(\tau_\pi)^2}\right]},
\end{equation}
the average $w$-component of the Bloch vector is calculated,
\begin{eqnarray}\label{e:w(2TauPi)Integration}
w_{\rm echo,hom}(2\tau_\pi)&=&\int_{-\infty}^\infty
-\cos\left(\Delta\delta\,\tau_\pi\right)\,
p_{\tau_\pi}(\Delta\delta)\, d\Delta\delta\nonumber\\
&=&\exp\left[-\tfrac{1}{2}\tau_\pi^2\sigma(\tau_\pi)^2\right].
\end{eqnarray}
Thus, the spin-echo visibility, $V$, yields
\begin{equation}\label{e:FitContrastOfSpinEchoTau}
    V(2\tau_\pi) = V_0 \,
  \exp{\left[-\tfrac{1}{2}\tau_\pi^2\,\sigma(\tau_\pi)^2\right]}.
\end{equation}

For comparison with the experimental values, we fit the spin echo
visibility of Fig.~\ref{fig:ContrastOfSpinEcho} with a Gaussian,
\begin{equation}\label{e:FitContrastOfSpinEcho}
    V(2\tau_\pi) = C_0 \exp{\left[-\frac{1}{2}\tau_\pi^2\sigma_{\rm
    exp}^2\right]}
\end{equation}
with a time-independent detuning fluctuation $\sigma_{\rm exp}$.
We define the homogeneous dephasing time $T_2\pe$ as the $1/e$
decay time of the spin echo visibility:
\begin{equation}\label{e:DefT2Prime}
    V(2T_2\pe) =C_0e^{-1}\quad\Rightarrow\quad T_2\pe =
    \frac{\sqrt{2}}{\sigma_{\rm exp}}.
\end{equation}

\subsection{Origins of irreversible dephasing}\label{subsec:originsOfDephasing}
Candidates for irreversible dephasing mechanisms include intensity
fluctuations (1) and pointing instability of the dipole trap laser
(2), heating of the atoms (3), fluctuating magnetic fields (4),
fluctuations of the microwave power and pulse duration (5) and
spin relaxation due to spontaneous Raman scattering from the
dipole trap laser (6). \\

\newlength{\tw}
\setlength{\tw}{2.5cm}
\begingroup
\squeezetable
\begin{table}[!h]
\begin{center}
\begin{tabular}{|c|c|rcrl|rcrl|rcrl|}
\cline{2-14}
  \multicolumn{1}{c|}{}
 &  \multicolumn{1}{c|}{$U_0$}

  & \multicolumn{4}{|c|}{1.0 mK}
  & \multicolumn{4}{|c|}{0.1~mK}
  & \multicolumn{4}{|c|}{0.04~mK} \\
\cline{2-14} \cline{2-14}
 \multicolumn{1}{c|}{}
 \vspace{-0.25cm}&
&&&&&&&&&&&&\\

\multicolumn{1}{c|}{} & $\sigma_{\rm exp}$ (meas.)
   &  22.0 &$\pm$&0.9&Hz
   & \tabentry{6.6}{0.2}{Hz}
   & \tabentry{1.54}{0.07}{Hz}
    \\
 \multicolumn{1}{c|}{}\vspace{-0.25cm}&&&&&&&&&&&&&\\
\hline
\vspace{-0.25cm}&&&&&&&&&&&&&\\
 (1)&\parbox{\tw}{intensity fluctuations\vspace{0,5mm}}
   & \ctabentry{$5.9$}{Hz}
   & \ctabentry{$0.67$}{Hz}
   & \ctabentry{$0.17$}{Hz}
    \\
\vspace{-0.25cm}&&&&&&&&&&&&&\\

\hline

&\parbox{\tw}{ pointing instability}&&&&&&&&&&&&\vspace{-0,5mm}\\
(2)&
   best case & \ctabentry{$10.6$}{Hz}
   & \ctabentry{$2.4$}{Hz}
   & \ctabentry{$1.3$}{Hz}
   \vspace{-1mm}
    \\
& worst case
   & \ctabentry{$21.6$}{Hz}
   & \ctabentry{$6.7$}{Hz}
   & \ctabentry{$3.7$}{Hz}
    \\
\vspace{-0.25cm}&&&&&&&&&&&&&\\
\hline
\vspace{-0.25cm}&&&&&&&&&&&&&\\
(3a)&\parbox{\tw}{heating\\$\sigma^{\rm(3)}_{\rm h}/2\pi$\\(upper
limit) }
   & \ctabentry{5.3}{Hz}
   & \ctabentry{1.6}{Hz}
   & \ctabentry{2.0}{Hz}\\
\vspace{-0.25cm}&&&&&&&&&&&&&\\
\vspace{-0.25cm}&&&&&&&&&&&&&\\
(3b)&\parbox{\tw}{photon scattering\\$\sigma_{\rm p}(T_2')/2\pi$}
   & \ctabentry{4.5}{Hz}
   & \ctabentry{1.5}{Hz}
   & \ctabentry{1.4}{Hz}\\

\vspace{-0.25cm}&&&&&&&&&&&&&\\

\hline
\vspace{-0.25cm}&&&&&&&&&&&&&\\

(4)&\parbox{\tw}{magnetic field fluctuations\\$\sigma_{\rm
b}(T_2\pe)/2\pi$}
   & \ctabentry{1.7}{Hz}
   & \ctabentry{0.35}{Hz}
   & \ctabentry{0.17}{Hz}

\\
\vspace{-0.25cm}&&&&&&&&&&&&&\\
  \hline
 \end{tabular}
\end{center}
\vspace{-0.5cm} \caption{Summary of dephasing mechanisms. Shown
are the fluctuation amplitudes $\sigma(T_2')/2\pi$.}
\label{tab:SummaryDephasingMechanisms}
\end{table}
\endgroup

\begin{figure}
\begin{center}
  \includegraphics[width=6cm]{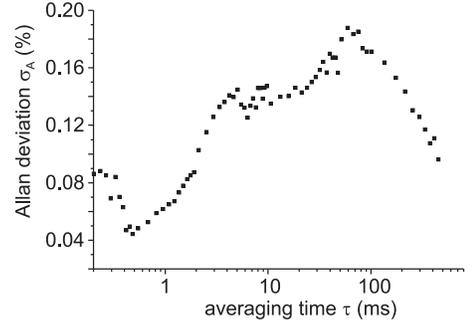}
\end{center}
\vspace{-0.5cm}\caption{
   Allan deviation of the intensity fluctuations
  according to Eq.~(\ref{e:AllanVar}).} \label{fig:IntensityFluct}
\end{figure}

{\it (1) Intensity fluctuations of the trapping laser.} The
intensity fluctuations are measured by shining the laser onto a
photodiode and recording the resulting voltage as a function of
time. From this signal we calculate $\sigma(\tau_\pi)^2$ by means
of the Allan variance, defined as \cite{Allan66}:
\begin{equation}\label{e:AllanVar}
\sigma_{\rm A}^2(\tau)
 =
 \frac{1}{m}\sum_{k=1}^m
 \frac{(\overline{x}_{\tau,k+1}-\overline{x}_{\tau,k})^2}{2}.
\end{equation}
Here $\overline{x}_{\tau, k}$ denotes the average of the
photodiode voltages over the $k$-th time interval $\tau$,
normalized to the mean voltage of the entire dataset. The
resulting Allan deviation $\sigma_{\rm A}$  is a dimensionless
number which expresses the relative fluctuations. They directly
translate into fluctuations $\sigma(\tau)$ of the detuning,
\begin{equation}\label{e:sigma(tau)}
\sigma(\tau)=\sqrt{2}\delta_0\sigma_{\rm A}(\tau).
\end{equation}
The factor of $\sqrt{2}$ arises because $\sigma(\tau)$ is the
standard deviation of the difference of two detunings with
standard deviation $\sigma_{\rm A}(\tau)$ each. The maximum
differential light shift $\delta_0$ in Eq.~(\ref{e:sigma(tau)}) is
calculated according to Eq.~(\ref{e:delta0FromFit}) using the
measured values of $\delta\pe$. As a result we find relative
intensity fluctuations of $\sigma_{\rm A}(\tau) < 0.2\%$ (see
Fig.~\ref{fig:IntensityFluct}). The corresponding absolute
fluctuation amplitudes $\sigma(T_2\pe)/2\pi$ (shown in
Table~\ref{tab:SummaryDephasingMechanisms}) are too weak to
account for the observed decay of the spin echo visibility.\\

\begin{figure}
\begin{center}
  \includegraphics[width=6cm]{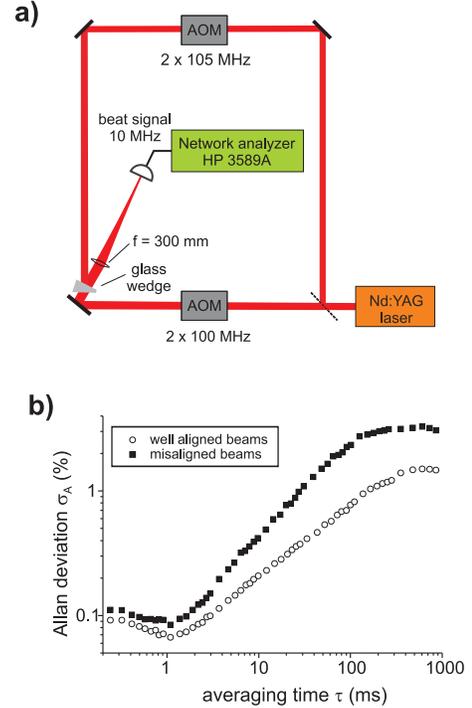}
\end{center}
\vspace{-0.5cm}\caption{Measuring the pointing instability. (a)
The dipole trap beams having a frequency difference of
$\Delta\nu=10$~MHz are overlapped on a fast photodiode. (b) Allan
deviation of the amplitude of the resulting beat signal.}
  \label{fig:PointingInstabilities}
\end{figure}

{\it (2) Pointing instability of the trapping laser.}  Any change
of the relative position of the two interfering laser beams also
changes the interference contrast, and hence the light shift
$\delta_0$. These position shifts can arise due to shifts of the
laser beam itself, due to variations of the optical paths
e.\,g.~from acoustic vibrations of the mirrors or from air flow.
In order to measure the pointing instabilities we mutually detune
the two dipole trap beams by $\Delta\nu=10$~MHz using the AOMs and
overlap them on a fast photodiode (see
Fig.~\ref{fig:PointingInstabilities}(a)). The amplitude of the
resulting beat signal directly measures the interference contrast
of the two beams and is thus proportional to the depth of the
potential wells of the standing wave dipole trap. We used a
network analyzer (HP~3589A) operated in ``zero span'' mode to
record the temporal variation of the beat signal amplitude within
a filter bandwidth of 10~kHz.

The resulting Allan deviation of the beat signal amplitudes is
shown in Fig.~\ref{fig:PointingInstabilities}(b). The lower curve
shows the signal in the case of well overlapped beams, whereas for
the upper curve, we purposely misaligned the beams so that the
beat signal amplitude is reduced by a factor of 2. In the latter
case variations of the relative beam position cause a larger
variation of the beat signal amplitude, since the beams overlap on
the slopes of the Gaussian profile.

These two curves measure the best and the worst cases of the
fluctuations. We found that the relative fluctuations for long
time scales of $\tau>100$~ms reach up to 3\% in the worst case.
They are thus one order of magnitude greater than the variations
caused by intensity fluctuations. The frequency fluctuations
$\sigma(\tau)$ are again calculated using
Eq.~(\ref{e:sigma(tau)}). This result is plotted together with the
observed visibility in Fig.~\ref{fig:ContrastOfSpinEcho}. Our data
points lie in between these best and worst case predictions.\\

{\it (3a) Heating effects.} Heating processes in the trap can also
cause significant irreversible decoherence, since they cause a
variation of the atomic resonance frequency within the microwave
pulse sequence. A constant heating rate $\dot{E}$ increases the
average energy of the atoms for the second free precession
interval $[\tau_\pi, 2\tau_\pi]$, compared to the first interval
$[0,\tau_\pi]$, by $\dot{E}\tau_\pi$. The energy $E$ of individual
atoms, however, can be changed by much more than this average
energy gain.

To estimate the effect we have to calculate the typical energy
change of individual atoms during the free precession interval
caused by the fluctuating forces which are responsible for the
heating. For this purpose we approximate the trap as harmonic and
assume the following model of the heating process. Due to a random
walk in momentum space, an initial atomic momentum $p=\sqrt{2mE}$
evolves into a symmetric, Gaussian momentum distribution around
$p$ with a standard deviation of $\Delta p_{\rm rms}$ given by the
average energy gain $\dot{E}\tau_\pi$
\begin{equation}
    \dot{E}\tau_\pi = \frac{(\Delta p_{\rm rms})^2}{2m}.
\end{equation}
Assuming $E \gg \dot{E}\tau_\pi$ we can linearly approximate the
energy-momentum relationship at $E$. In this approximation the
momentum distribution is therefore equivalent to a Gaussian
distribution of the energies with
\begin{equation}\label{Erms}
    \Delta E_{\rm rms} = 2\sqrt{E \dot{E} \tau_\pi}.
\end{equation}
According to Eq. (\ref{e:deltaLS}) the corresponding standard
deviation $\sigma_{{\rm heat},E}$ of the detunings $\Delta\delta$
is
\begin{equation}\label{DDRMS}
    \sigma_{{\rm heat},E}(\tau_\pi) =\frac{\eta}{\hbar} \sqrt{E \dot{E}
    \tau_\pi} ,
\end{equation}
depending on the initial energy E.

We now integrate the distribution of the detunings
\begin{eqnarray}\label{pdeltaDeltaE}
&&\!\!\!\!\!\!\!\!\!\!\!    p_E(\Delta\delta) =\nonumber\\
 &&  \!\!\!\!\!\! =  \frac{1}{\sqrt{2 \pi}}\frac{1} {\sigma_{{\rm heat},E}(\tau_\pi)}
 \exp{\left(-\frac{\Delta\delta^2}{2 \bigl( \sigma_{{\rm heat},E}(\tau_\pi) \bigr)^2}\right)}
\end{eqnarray}
over the initial energy $E$ weighted by the $n$-dimensional
thermal energy distribution $p^{(n)}(E) \propto E^{n-1}
\exp{\left(-{E}/{k_{B}T}\right)}$:
\begin{equation}\label{pII_n}
    p^{(n)}(\Delta\delta) = \int_0^{\infty} p_E(\Delta\delta) p^{(n)}(E) dE.
\end{equation}
Finally we obtain the rms detuning fluctuations $\sigma_{\rm
heat}$ from the resulting distribution of $\Delta\delta$ as
\begin{equation}
\left[ \sigma^{(n)}_{\rm heat}(\tau_\pi)\right]^2 =
\int_{-\infty}^{\infty} \Delta\delta^2 p^{(n)}
  (\Delta\delta) d\Delta\delta .
\end{equation}
Evaluation of $\sigma^{(n)}_{\rm heat}$ for the experimentally
relevant time scale $\tau_\pi=T_2'/2$ yields
\begin{equation}\label{sigmaHEND}
    \sigma^{(n)}_{\rm heat} = \frac{\eta}{\hbar} \sqrt{\frac{n}{2} \dot E T_2' k_B T} .
\end{equation}
Heating effects in our trap have been investigated in detail in
Ref.~\cite{Alt02b}. An upper limit for the heating rate of $\dot E
= 2\cdot 10^{-2}$~mK/s is obtained from the trap lifetime of 50~s
(for $U_0 = 1.0$~mK). When we linearly scale $\dot E$ to our trap
depths of $U_0 = 1.0$~mK, $U_0 = 0.1$~mK, and $U_0 = 0.04$~mK, and
assume temperatures of $T = 0.1$~mK, $T =0.06$~mK, and $T
=0.02$~mK, we obtain fluctuation amplitudes for the 3D-case
($n=3$) of $\sigma^{(3)}_{\rm heat}=5.3$~Hz, $\sigma^{(3)}_{\rm
heat}=1.6$~Hz and $\sigma^{(3)}_{\rm heat}=2.0$~Hz, respectively.
We stress however, that these values for $\sigma_{\rm heat}$ are
upper limits since we did not measure heating rates $\dot E$ for
the trap depths we used. The actual values of $\dot E$ and the
resulting values for $\sigma_{\rm heat}$ could be orders of
magnitude smaller than the upper limits inferred from the life
time because the heating rate strongly depends on the oscillation
frequencies and the details of the laser
noise spectrum.\\

{\it (3b) Photon recoil.} Our model of the heating process also
gives an estimate of the dephasing due to photon recoil. If we had
one photon scattered per time interval $\tau_\pi$ giving two
recoils, we would obtain a heating rate
\begin{equation}\label{Edot}
    \dot E = \frac{\hbar^2 k^2}{m}\frac{1}{\tau_\pi}.
\end{equation}
Inserting into Eq.~(\ref{sigmaHEND}) ($n=3$) yields
\begin{equation}\label{sigmaH1D1PH}
    \sigma_{\rm heat}^{\rm 1 ph} = \eta k \sqrt {\frac{3 k_B T}{m}} .
\end{equation}
Scattering of $n_{\rm ph}$ photons would yield
\begin{equation}\label{sigmaH1D1PHn}
    \sigma_{\rm ph}(n_{\rm ph}) = \sqrt{n_{\rm ph}} \,\sigma_{\rm heat}^{\rm 1 ph}.
\end{equation}

Given a scattering rate $\Gamma_{\rm s}$, the number of scattered
photons obeys a Poissonian distribution. Since for our parameters,
the probability of scattering more than one photon is negligible,
we obtain
\begin{equation}\label{deDZ}
    \sigma_{\rm ph}(\tau_\pi) =  \eta k \sqrt {\frac{3 k_B T \Gamma_{\rm s}\tau_\pi}{m}}
    \exp{\left(-\frac{\Gamma_{\rm s} \tau_\pi}{2} \right)} .
\end{equation}
We use the temperatures of the previous paragraph and the photon
scattering rates (see below) of $\Gamma_{\rm s} =10.6$~s$^{-1}$,
$\Gamma_{\rm s} = 1.06$~s$^{-1}$ and $\Gamma_{\rm s}
=0.41$~s$^{-1}$. With $\tau_\pi=T_2'/2$ we obtain $\sigma_{\rm
ph}(T_2')=4.5$~Hz, $\sigma_{\rm ph}(T_2')=1.5$~Hz and $\sigma_{\rm ph}(T_2')=1.4$~Hz,
respectively.\\

 {\it (4) Fluctuating magnetic fields.} Using a fluxgate
magnetometer we measured a peak-to-peak value of the magnetic
field fluctuations of $\Delta B=0.13~\mu$T, dominated by
components at $\nu=50$~Hz. The resulting frequency shift on the
\ket{F=4,m_F=0} $\rightarrow$ \ket{F=3,m_F=0} transition is:
\begin{equation}\label{e:BField}
    \Delta \omega=2\,\Delta\omega_{0\rightarrow0}\,B_0\, \Delta B,
\end{equation}
where \mbox{$B_0=97.9~\mu$T} is the offset field and
$\Delta\omega_{0\rightarrow0}/2\pi = 43$~mHz/($\mu$T)$^2$ is the
quadratic Zeeman shift. For our case, we obtain
$\Delta\omega=1.1$~Hz.

The effect of the magnetic fluctuations depends on the time
interval between the microwave pulses. If this time is large
compared to $1/\nu$, all fluctuations cancel except for those of
the last oscillation period. As a consequence, the effect on the
detuning fluctuations $\sigma$ also decreases. We calculate this
effect by computing the Allan deviation $\sigma_{\rm A}(\tau)$ of
a 50~Hz sine signal. The detuning fluctuations then read
$\sigma_{\rm b}(\tau)=\sqrt{2}\Delta \omega\,\sigma_{\rm
A}(\tau)$. The resulting $\sigma_{\rm b}(T_2\pe)$, shown in
Table~\ref{tab:SummaryDephasingMechanisms}, is too small to
account for the decay of the spin echo amplitude.\\

{\it (5) Fluctuation of microwave power and pulse durations.} The
application of two $\pi/2$-pulses and one $\pi$-pulse results in
 $w_{\rm echo}(2\tau_\pi)=-1$. Any fluctuations of the amplitude
($\Delta\Omega_{\rm R}/\Omega_{\rm R}$) or pulse duration ($\Delta
\tau/\tau$) result in  variations of the amplitude of the spin
echo signal, i.\,e.~$w_{\rm echo}(2\tau_\pi)=-\cos\Delta\phi$
according to:
\begin{equation}\
 \left(\frac{\Delta\phi}{2\pi}\right)^2  =
  \left(\frac{\Delta\Omega_{\rm R}}{\Omega_{\rm R}}\right)^2
  +
  \left(\frac{\Delta \tau}{\tau}\right)^2
\end{equation}
With $\Delta \tau/\tau < 10^{-3}$ (measured) and
$\Delta\Omega_{\rm R}/\Omega_{\rm R}<10^{-2}$ (specifications of
the synthesizer) we obtain $\Delta\phi/2\pi < 10^{-2}$, which is
too small to be observed. Moreover, this effect neither depends on
the dipole trap depth nor on the time delay between the microwave
pulses.

The timing of the microwave pulses would be affected by a clock
inaccuracy of the D/A-board of the computer control system which
triggers the microwave pulses. Its specified accuracy
$\Delta\tau/\tau=10^{-4}$, results in a phase fluctuation
$\delta\pe\tau_\pi \,\Delta\tau/\tau < 0.01$ for all parameters
$\delta\pe$ and $\tau_\pi$ used in our experiment. Thus, the
fluctuations of microwave power, pulse duration, and timing do not
account for the observed reduction of the spin echo visibility.
\\

{\it (6) Spin relaxation due to light scattering.} The population
decay time, $T_1$, is governed by the scattering of photons from
the dipole trap laser, which couples the two hyperfine ground
states via a two-photon Raman transition.
In our case, the hyperfine changing spontaneous Raman processes
are strongly reduced due to a destructive interference of the
transition amplitudes. Thus, the spin relaxation rate is much
larger than the spontaneous scattering rate. This effect was first
observed on optically trapped Rubidium atoms in the group of
D.~Heinzen \cite{Cline94} and was also verified in experiments in
our group \cite{Frese00}.

The corresponding transition rate is calculated by means of the
Kramers-Heisenberg formula \cite{Loudon}, which is a result from
second order perturbation theory. We obtain for the rate of
spontaneous transitions, $\Gamma_{\rm s}$, from the ground state
$\ket{F,m}$ to the ground state $\ket{F'',m''}$:
\begin{equation}\label{e:KramersHeisenberg}
\Gamma_{\rm s}
 =
\frac{3c^2\omega_{\rm L}^3I}{4\hbar\, d^4}
\left|
 \frac{a^{\mbox{\tiny (1/2)}}}{\Delta_{\rm 1/2}}+
 \frac{a^{\mbox{\tiny (3/2)}}}{\Delta_{\rm 3/2}}
\right|^2,
\end{equation}
where $\Delta_{\rm J'}=\omega_{\rm L}-\omega_{\rm J'}$ is the
detuning of the dipole trap laser from the $^6\!P_{\rm J'}$-state
and $d=\bra{4,4}\mu_{+1}\ket{5,5}$, with the dipole operator
$\mu_{+1}$ for $\Delta m = +1$ transitions. The transition
amplitudes $a^{\mbox{\tiny (J')}}$ are obtained by summing over
all possible intermediate states \ket{F',m'} of the relevant
$^6\!P_{\rm J'}$ manifold \cite{Cline94}. For Rayleigh scattering
processes, which do not change the hyperfine state ($F,M=F'',
M''$), the amplitudes add up, $a^{\mbox{\tiny
(3/2)}}=2a^{\mbox{\tiny (1/2)}}$. However, for state changing
Raman processes ($F,M \neq F'', M''$), the two transition
amplitudes are equal but have opposite sign, $a^{\mbox{\tiny
(3/2)}}=-a^{\mbox{\tiny (1/2)}}$. Then the two terms in
Eq.~(\ref{e:KramersHeisenberg}) almost cancel in the case of far
detuning, $\Delta_{1/2}\approx\Delta_{3/2}$. As a result the
spontaneous Raman scattering rate scales as $1/\Delta^4$ whereas
the Rayleigh scattering rate scales as $1/\Delta^2$. The
suppression factor can be expressed using the fine structure
splitting $\Delta_{\rm fs}=\Delta_{\rm 3/2}-\Delta_{\rm 1/2}$ as
\begin{equation}\label{e:RamanVSRayleigh}
 \Gamma_{\rm Raman}
 =
 \beta\,\Gamma_{\rm s}
 \qquad \mbox{with} \quad
 \beta =
 \left|
    \frac{\Delta_{\rm fs}}{3\Delta_{\rm 1/2}}
 \right|^2.
\end{equation}
For the case of cesium, we obtain a suppression factor of $\beta =
0.011$. The Rayleigh scattering rate for an atom trapped in a
potential of $U_0=1.0$~mK is $\Gamma_{\rm s}=11$~s$^{-1}$. Then,
the corresponding spontaneous Raman scattering rate is
$\Gamma_{\rm Raman}=0.12$~s$^{-1}$ and the population decay time
$T_1 = \Gamma_{\rm Raman}^{-1} = 8.6$~s. Since in most of our
experiments, the trap depth is significantly smaller, $T_1$ will
be even larger. As a consequence, we neglect the population decay
due to spontaneous scattering. Note that the experiments of Refs.
\cite{Cline94,Frese00} were only sensitive to changes of the
hyperfine $F$-state, since the atoms were in a mixture of
$m_F$-sublevels. However, the theoretical treatment above predicts
similarly long relaxation times for any particular $m_F$-sublevel,
which is consistent with our observations. \\

\subsection{Conclusions}
We have developed an analytical model which treats the various
decay mechanisms of the hyperfine coherence of trapped cesium
atoms independently. This is justified by the very different time
scales of the decay mechanisms ($T_2^*\ll T_2\pe\ll T_1$). Our
model reproduces the observed shapes of Ramsey and spin echo
signals, whose envelopes are the Fourier transform of the energy
distribution of the atoms in the trap.

The irreversible decoherence rates manifest themselves in the
decay of the spin echo visibility and are caused by fluctuations
of the atomic resonance frequency in between the microwave pulses.
In the above analysis we have investigated various dephasing
mechanisms and characterized them by the corresponding amplitude
of the detuning fluctuations which are summarized in
Table~\ref{tab:SummaryDephasingMechanisms}. We find that a major
mechanism of irreversible dephasing is the pointing instability of
the dipole trap laser beams resulting in fluctuations of the trap
depth and thus the differential light shift. Significant
decoherence is also caused in the shallow dipole trap by heating
due to photon scattering. Heating due to technical origin, such as
fluctuations of the depth and the position of the trap, cannot be
excluded as an additional
source of decoherence.\\

Compared to our experiment, significantly longer coherence times
($T_2^*=4$\,s) were observed by N.~Davidson and S.~Chu in blue
detuned traps in which the atoms are trapped at the minimum of
electric fields \cite{Davidson95}. In Ref.~\cite{Davidson95},
$T_2^*=15$~ms obtained with sodium atoms in a Nd:YAG dipole trap
($U_0=0.4$~mK) was reported, which is comparable to our
observation. In other experiments, the inhomogeneous broadening
has been reduced by the addition of a weak light field, spatially
overlapped with the trapping laser field and whose frequency is
tuned in between the two hyperfine levels \cite{Kaplan02}. Of
course, cooling the atoms to the lowest vibrational level by using
e.\,g.~Raman sideband cooling techniques \cite{Hamann98,Perrin98},
would also reduce inhomogeneous broadening. The  magnetic field
fluctuations could possibly be largely suppressed by triggering
the experiment to the 50~Hz of the power line.\\

\begin{acknowledgements}
We thank Nir Davidson for valuable discussions. This work was
supported by the Deutsche Forschungsgemeinschaft and the EC.
\end{acknowledgements}

\end{document}